\documentclass[a4paper,11pt]{article}
\pdfoutput=1
\usepackage{jheppub}
\usepackage[T1]{fontenc}
\usepackage{amsmath}
\usepackage{mathtools}
\DeclareMathOperator{\re}{Re}
\DeclareMathOperator{\dd}{d}

\newcommand{\SU}{\mathrm{SU}}
\newcommand{\U}{\mathrm{U}}
\newcommand{\Orth}{\mathrm{O}}
\newcommand{\Z}{\mathbb{Z}}
\newcommand{\Tc}{T_{\mbox{\tiny{c}}}}
\newcommand{\SW}{S_{\mbox{\tiny{W}}}}
\newcommand{\mD}{m_{\mbox{\tiny{D}}}}
\newcommand{\TKT}{T_{\mbox{\tiny{KT}}}}
\newcommand{\Ploop}{\mathcal{P}}
\newcommand{\el}{W}
\newcommand{\El}{E}
\newcommand{\Ecore}{E_{\mbox{\tiny{{c}}}}}
\newcommand{\rhos}{\rho_{\mbox{\tiny{{s}}}}^{\mbox{\tiny{{r}}}}}
\newcommand{\Xc}{\mathcal{X}_{\mbox{\tiny{{c}}}}}
\newcommand{\KR}{K_{\mbox{\tiny{{r}}}}}
\newcommand{\Kc}{K_{\mbox{\tiny{{c}}}}}
\newcommand{\yR}{y_{\mbox{\tiny{{r}}}}}
\newcommand{\nf}{n_{\mbox{\tiny{{f}}}}}
\newcommand{\psiin}{\psi_{\mbox{\tiny{{in}}}}}
\newcommand{\betac}{\beta_{\mbox{\tiny{{c}}}}}
\newcommand{\redchisq}{\chi^2_{\mbox{\tiny{{red}}}}}

\begin{document}
\begin{titlepage}
\begin{flushright} 
DESY-19-027
\end{flushright}
\vskip0.5cm
\renewcommand\thefootnote{\mbox{$\fnsymbol{footnote}$}}
\begin{center}
{\Large\bf Conformal field theory and the hot phase of three-dimensional $\U(1)$ gauge theory}
\end{center}
\vskip1.3cm
\centerline{Michele~Caselle$^{a,b,c}$, Alessandro~Nada$^{d}$, Marco~Panero$^{a,c}$, and Davide~Vadacchino$^{e}$}
\vskip1.5cm
\centerline{\sl $^a$Department of Physics and $^b$Arnold-Regge Center, University of Turin, and $^c$INFN, Turin}
\centerline{\sl Via Pietro Giuria 1, I-10125 Turin, Italy}
\vskip0.5cm
\centerline{\sl $^d$John von Neumann Institute for Computing, DESY}
\centerline{\sl Platanenallee 6, D-15738 Zeuthen, Germany}
\vskip0.5cm
\centerline{\sl $^e$INFN, Pisa}
\centerline{\sl Largo Bruno Pontecorvo 3, I-56127 Pisa, Italy}
\vskip0.5cm
\begin{center}
{\sl  E-mail:} \hskip 1mm \href{mailto:caselle@to.infn.it}{{\tt caselle@to.infn.it}}, \href{mailto:alessandro.nada@desy.de}{{\tt alessandro.nada@desy.de}},\\ \href{mailto:marco.panero@unito.it}{{\tt marco.panero@unito.it}}, \href{mailto:davide.vadacchino@pi.infn.it}{{\tt davide.vadacchino@pi.infn.it}}
\end{center}
\vskip1.0cm
\begin{abstract}

We study the high-temperature phase of compact $\U(1)$ gauge theory in $2+1$ dimensions, comparing the results of lattice calculations with analytical predictions from the conformal-field-theory description of the low-temperature phase of the bidimensional XY model. We focus on the two-point correlation functions of probe charges and the field-strength operator, finding excellent quantitative agreement with the functional form and the continuously varying critical indices predicted by conformal field theory.
\end{abstract}

\end{titlepage}

\section{Introduction}
\label{sec:introduction}

Quantum electrodynamics (QED) in three spacetime dimensions is an interesting theoretical model, with many applications relevant for high- and low-energy physics. On the one hand, it shares important qualitative features such as charge confinement and dynamical chiral-symmetry breaking with non-Abelian gauge theories in four dimensions~\cite{Fiebig:1990uh}. On the other hand, it also provides a useful effective description of the long-wavelength physics for different condensed-matter systems~\cite{Hosotani:1977cp, Baskaran:1987my, Fradkin:1991nr, Frohlich:1993gs, Diamantini:1995yb, Franz:2002qy, Herbut:2002yq, Senthil:2004aza, Lee:2006zzc, Gusynin:2007ix}.

Thanks to its relative mathematical simplicity, this is one of the few quantum field theories in which non-trivial dynamical properties can be studied analytically. Classical results include the seminal studies by Polyakov~\cite{Polyakov:1975rs, Polyakov:1976fu}: his semiclassical calculations showed that the ground state of the theory is a plasma\footnote{Note that the finiteness of the screening length for a Coulomb gas in three spacetime dimensions can be proven by renormalization-group arguments~\cite{Kosterlitz:1977tdd}.} of monopoles (which are instanton-like objects in three dimensions), leading to a linearly confining potential for static electric probe charges and to a finite mass gap, for all positive values of the gauge coupling $e$. In this setup, a crucial ingredient for the existence of monopoles is the compactness of the $\U(1)$ gauge group, which is realized when the theory is regularized on a lattice~\cite{Wilson:1974sk} or when $\U(1)$ is a subgroup of a compact group, as, for example, in the Georgi-Glashow model~\cite{Georgi:1972cj}. Another milestone in the literature on this theory was the analytical proof, due to G\"opfert and Mack~\cite{Gopfert:1981er}, of the existence of a non-zero mass gap and of a finite string tension, in the Villain formulation of the model~\cite{Villain:1974ir}.

Other analytical studies have investigated the interplay of topological properties in three-dimensional spacetime and the generation of mass for gauge fields~\cite{Deser:1981wh}, the structure of perturbative expansions for this super-renormalizable theory~\cite{Jackiw:1980kv}, chiral-symmetry breaking~\cite{Pisarski:1984dj, Appelquist:1985vf, Nash:1989xx, Liu:2002yc, Kaveh:2004qa} and the qualitative change in vacuum structure driven by a sufficiently large number of dynamical fermion species~\cite{Appelquist:1988sr, Azcoiti:1993ey, Kubota:2001kk, Kleinert:2002uv, Herbut:2003bs, Appelquist:2004ib, Fischer:2004nq, Nogueira:2005aj, Nogueira:2007pn, Braun:2014wja, Huh:2014eea, DiPietro:2015taa, Janssen:2016nrm, Giombi:2015haa, Giombi:2016fct, Herbut:2016ide, Chester:2016wrc, Chester:2016ref, Kotikov:2016prf, Gusynin:2016som, Thomson:2016ttt, Thomson:2017dut, DiPietro:2017kcd, DiPietro:2017vsp, Benvenuti:2018cwd, Li:2018lyb, Steinberg:2019uqb, Benvenuti:2019ujm}, and a number of other interesting aspects~\cite{Parga:1981tm, Heller:1981bk, Marston:1990bj, Aitchison:1992ik, Cangemi:1994by, Kogan:1994vb, Brown:1997gk, Kovner:1998eg, Antonov:1998kw, Mavromatos:1999jf, Onemli:2001nf, Agasian:2001an, Fosco:2005ae, Unsal:2008sc, Wang:2010in, ElShowk:2011gz, Cherman:2017dwt, Diamantini:2018jyj, Maggiore:2019wie}.

In parallel with these analytical studies, QED in three spacetime dimensions has also been extensively investigated by means of lattice simulations: this has been done both with~\cite{Dagotto:1988id, Dagotto:1989td, Fiebig:1990uh, Hands:2002dv, Hands:2004bh, Hands:2004ex, Hands:2006dh, Armour:2011zx, Raviv:2014xna, Karthik:2015sgq, Karthik:2016ppr, Xu:2018wyg} and without~\cite{DeGrand:1980eq, Sterling:1983fs, Karliner:1983ab, Coddington:1986jk, Wensley:1989ja, Trottier:1993nf, Baig:1994ia, Chernodub:2001ws, Chernodub:2001da, Chernodub:2001mg, Chernodub:2002gp, Loan:2002ej, Chernodub:2003bb, Arakawa:2005nc, Fiore:2005ku, Fiore:2005ps, Fiore:2008kq, Borisenko:2008sc, Borisenko:2010qe, Borisenko:2013jwa, Borisenko:2015jea, Caselle:2014eka, Caselle:2016mqu, Chernodub:2017mhi, Chernodub:2017gwe, Athenodorou:2018sab} dynamical fermion fields.

The behavior of $\U(1)$ gauge theory (regularized on the lattice) at finite temperature $T$ and without matter fields, which has been studied in refs.~\cite{Parga:1981tm, Coddington:1986jk, Chernodub:2001ws, Chernodub:2001da, Chernodub:2001mg, Chernodub:2002gp, Borisenko:2008sc, Borisenko:2010qe, Borisenko:2013jwa, Borisenko:2015jea}, is particularly interesting: there exists a finite critical temperature $\Tc$ such that linear confinement persists for temperatures $T < \Tc$, whereas for $T>\Tc$ the potential $V$ associated with a pair of static charges grows logarithmically with their spatial separation $r$. This can be compared and contrasted with what happens in $\SU(N)$ gauge theories in $3+1$ spacetime dimensions~\cite{Svetitsky:1985ye}, which exhibit a linearly confining phase at low temperatures and a phase transition to a deconfined phase at a finite temperature. This deconfinement transition can be interpreted in terms of spontaneous breakdown of a global symmetry based on the center $\Z_N$ of the gauge group: the order parameter is the average Polyakov loop $\Ploop$, i.e. the trace of a Wilson line winding around the Euclidean-time direction~\cite{Kuti:1980gh, McLerran:1980pk, McLerran:1981pb}. After renormalization of an ultraviolet divergence~\cite{Dotsenko:1979wb} (see also ref.~\cite{Mykkanen:2012ri} and references therein), the average Polyakov loop can be directly related to the free energy associated with a chromoelectric probe charge: in the thermodynamic limit $\langle \Ploop \rangle$ vanishes for $T < \Tc$ (implying an infinite energy cost for the existence of an isolated fundamental color source in the confining phase, i.e. quark confinement), whereas it has a finite expectation value at $T > \Tc$. In contrast, the $\U(1)$ center symmetry of $\U(1)$ gauge theory in $2+1$ dimensions remains unbroken, and, while in the high-temperature phase the theory does not have a dynamically generated, finite, characteristic length scale, the logarithmic Coulomb potential is still sufficient to confine static charges.

As the finite-temperature transition in $\U(1)$ gauge theory in $2+1$ dimensions is continuous, one expects that at $T=\Tc$ the long-distance properties of the system are equivalent to those of a two-dimensional spin system with global $\U(1)$ symmetry~\cite{Svetitsky:1982gs}, i.e. the classical XY model, that exhibits a Kosterlitz-Thouless transition~\cite{Kosterlitz:1973xp} (see also refs.~\cite{Berezinsky:1970fr, Kosterlitz:1974sm, Jose:1977gm, Amit:1979ab}). In the past, the validity of this conjecture has been investigated in various numerical studies~\cite{Coddington:1986jk, Borisenko:2008sc, Borisenko:2010qe, Borisenko:2013jwa} and the most recent work gives conclusive evidence in support of it~\cite{Borisenko:2015jea}.

As discussed in ref.~\cite{Svetitsky:1982gs}, this correspondence relies on the continuous nature of the transition at $T=\Tc$. In turn, the existence of an infinite correlation length is also an essential necessary condition for scale and conformal invariance. In the two-dimensional XY model, this condition is realized in a peculiar way: even though the system can never have spontaneous magnetization~\cite{Mermin:1966fe}, at low temperatures the model is in a phase characterized by ``topological'' order~\cite{Kosterlitz:1973xp}, with two-point spin correlation functions decaying only with inverse powers of the spatial separation $r$ between the spins~\cite{McBryan:1977ot, Frohlich:1981yn}. The fact that the whole low-temperature phase of the two-dimensional XY model is gapless and admits a conformal-field-theory description raises the question, what happens in the corresponding phase of the three-dimensional gauge theory, i.e. the high-temperature phase? To answer this question, in this work we carry out a systematic study of compact $\U(1)$ lattice gauge theory at $T>\Tc$, and compare a large set of novel numerical results, obtained by Monte~Carlo simulations, with analytical predictions derived from conformal field theory. Specifically, we focus our attention on correlation functions of plaquette operators, Polyakov loops, and on the profile of the flux tube induced by a pair of static probe charges.

Note that the approach we follow in the present work is different from the one of other studies, which analyzed the ``effective'' dimensional reduction of the XY model from three to two dimensions upon compactification of one of the system sizes~\cite{Janke:1993mc, Schultka:1994ze, Schultka:1996so, Schultka:1997be, Nho:2003hc, Zhang:2006fs, Hucht:2007tc, Vasilyev:2007mc, Hasenbusch:2009tk}.

The structure of the article is the following: in section~\ref{sec:u1}, we introduce the $\U(1)$ gauge theory in three dimensions, discussing its most important properties and its compact formulation on an isotropic cubic lattice. In section~\ref{sec:conformal_xy}, we present the conformal-field-theory predictions for the low-temperature phase of the two-dimensional XY model, and discuss their implications for the corresponding operators defined in the three-dimensional gauge theory. Our results are presented and analyzed in detail in section~\ref{sec:results}, while the final section~\ref{sec:conclusions} includes a summary of our findings, and a discussion of their implications. The appendix~\ref{app:RG_analysis_of_XY_model} presents a review of the renormalization-group analysis of the XY model. Throughout this article, we work in natural units, setting the speed of light in vacuum, the reduced Planck's constant, and Boltzmann's constant to unity.

\section{$\U(1)$ gauge theory in three spacetime dimensions}
\label{sec:u1}

The formulation of $\U(1)$ gauge theory (without matter fields) in three-dimensional continuum Minkowski spacetime is based on the action
\begin{equation}
\label{continuum_Minkowski_action}
S_{\mbox{\tiny{cont}}} = -\frac{1}{4} \int \dd^2x \int \dd t F_{\mu\nu}F^{\mu\nu},
\end{equation}
where the field strength is defined as $F_{\mu\nu}=\partial_\mu A_\nu - \partial_\nu A_\mu$; note that in three spacetime dimensions the gauge field $A$ and the electric charge $e$ have energy dimension $1/2$, the Coulomb potential is logarithmic, and the magnetic field is a scalar. The classical equations of motion derived from eq.~(\ref{continuum_Minkowski_action}) are $\partial_\mu F^{\mu\nu}=0$ and the definition of $F_{\mu\nu}$ implies that the Bianchi identity $\epsilon_{\mu\nu\rho} \partial^\mu F^{\nu\rho}=0$ is trivially satisfied. In turn, the latter property implies that one can reformulate the theory in terms of the free, massless scalar field $\phi$ such that $\partial_\mu \phi = \epsilon_{\mu\nu\rho} F^{\nu\rho}$.

At the quantum level, the most interesting physical properties of the theory become manifest when one studies it in its compact formulation, i.e. assuming the gauge field components $A_\mu$ to be periodic. If the theory is Wick-rotated to Euclidean spacetime and regularized on an isotropic cubic lattice $\Lambda$ of spacing $a$, one can introduce the link degrees of freedom $U_\mu(x)$ and the Wilson action~\cite{Wilson:1974sk}
\begin{equation}
\label{Wilson_action}
\SW = -\frac{1}{a e^2} \sum_{x \in \Lambda} \sum_{1 \le \mu < \nu \le 3} \re \, U_{\mu\nu}(x),
\end{equation}
where $U_{\mu\nu}(x) = U_\mu(x) U_\nu(x+a\hat{\mu})U_\mu^\star(x+a\hat{\nu})U_\nu^\star(x)$. For later convenience, we also introduce $\beta=1/(a e^2)$. The $U_\mu(x)$ variables are complex phases and can be thought of as parallel transporters relating the reference frames in internal space defined on two nearest-neighbor sites $x$ and $x+a\hat{\mu}$:
\begin{equation}
\label{link_definition}
U_\mu(x)=\exp\left[ i e a A_\mu \left(x+\frac{a}{2}\hat{\mu}\right) \right].
\end{equation}
Eq.~(\ref{link_definition}) makes it manifest that the theory defined by eq.~(\ref{Wilson_action}) is invariant under $A_\mu \to A_\mu + 2\pi k/(ea)$ for any integer $k$, i.e. that the gauge group is compact. The periodicity of the gauge field plays a crucial r\^ole in determining the long-wavelength properties of the theory: the gauge-field configurations admit topological defects, which can be thought of as ``magnetic monopoles'' (actually ``instantons'' of the theory defined in three spacetime dimensions). Their condensation in the ground state of the theory implies that the expectation value of the gauge holonomies of large contours decreases exponentially with the area they bound, i.e. confinement of electric charges~\cite{Polyakov:1976fu} as a dual Mei{\ss}ner effect~\cite{Nambu:1974zg, Mandelstam:1974pi, 'tHooft:1979uj}.

The calculations presented in ref.~\cite{Gopfert:1981er} show that, at large $\beta$, the mass gap $\mD$ and the string tension $\sigma$ characterizing the linearly confining potential of electric charges scale as
\begin{equation}
\label{semiclassical_mD}
\mD a \simeq k_1 \sqrt{\beta} \exp(-k_2 \beta)
\end{equation}
and
\begin{equation}
\label{semiclassical_sigma}
\sigma a^2 \simeq \frac{k_3}{\sqrt{\beta}} \exp(-k_2 \beta),
\end{equation}
where the numerical constants $k_1=2\pi\sqrt{2}$, $k_2 \simeq 0.2527 \pi^2$, and $k_3=4\sqrt{2}/\pi$ are evaluated in a semiclassical approximation, which is reliable for $\beta \gg 1$. 

Eqs.~(\ref{semiclassical_mD}) and~(\ref{semiclassical_sigma}) have interesting implications for the continuum limit of the lattice theory. In the ``na\"{\i}ve'' continuum limit ($a \to 0$ at fixed $e$) the screening length diverges and the string tension vanishes, so that the theory reduces to the continuum Maxwell theory of non-interacting photons~\cite{Gross:1984vg}. On the other hand, one can assume the continuum limit to be taken on a ``line of constant physics'' at fixed $\sigma$: then, given that at large $\beta$
\begin{equation}
\label{mD2_over_sigma}
\frac{\mD^2}{\sigma} \propto \sqrt{\beta^3} \exp(-k_2 \beta),
\end{equation}
$\mD$ tends to zero, namely the screening length diverges, and the continuum potential associated with a pair of probe charges is again purely Coulombic (i.e. logarithmic and unscreened) at all distances $r$. Conversely, if the continuum limit is taken at fixed $\mD$, then from eq.~(\ref{mD2_over_sigma}) it follows that $\sigma$ diverges: increasing the spatial separation $r$ between two probe charges by a finite amount $\Delta r$ would therefore require an infinite amount of energy, which means that it is not physically possible to couple charged matter fields to the theory. As a consequence, the theory never exhibits linear confinement in the continuum limit.

The lattice theories based on the Wilson action~\cite{Wilson:1974sk} defined in eq.~(\ref{Wilson_action}) and on the ``periodic Gau{\ss}ian'' action~\cite{Villain:1974ir} can be reformulated as a spin model~\cite{Banks:1977cc, Savit:1977fw, Glimm:1977gz, Baaquie:1977ey}: the Feynman path integral
\begin{equation}
\label{Z}
\mathcal{Z} = \int \prod_{x,\mu} \dd U_\mu(x) \exp[-\SW]
\end{equation}
(where $\dd U_\mu(x)$ is the Haar measure for $U_\mu(x)$) can be rewritten as the one for a lattice theory with integer-valued degrees of freedom $s$, defined on the sites of the dual lattice
\begin{equation}
\label{dual_Z}
\mathcal{Z} = \sum_{\{ s \}} \prod_{y,\nu} I_{|s(y)-s(y+a\hat{\nu})|} (\beta),
\end{equation}
where $I_\alpha(z)$ is the modified Bessel function of the first kind of order $\alpha$ and the product is taken over the bonds of the dual cubic lattice, with the appropriate boundary conditions. Similar relations hold for generic expectation values of gauge-invariant quantities $\mathcal{O}$
\begin{equation}
\label{generic_expectation_value}
\langle \mathcal{O} \rangle = \frac{1}{\mathcal{Z}} \int \prod_{x,\mu} \dd U_\mu(x) \mathcal{O} \exp[-\SW].
\end{equation}
In particular, the two-point correlation function $\langle P^\star (r) P(0) \rangle$ at separation $r = N_r a$ can be rewritten as
\begin{equation}
\label{dual_PP_correlator}
\langle P^\star (r) P(0) \rangle = \frac{\mathcal{Z}_{N_t \times N_r}}{\mathcal{Z}},
\end{equation}
having introduced
\begin{equation}
\label{Z_Nt_Nr}
\mathcal{Z}_{N_t \times N_r} = \sum_{\{s\}} \prod_{y,\nu} I_{|s(y)-s(y+a\hat{\nu})+ n_\nu(y)|} (\beta),
\end{equation}
where $n_\nu(y)$ vanishes on all oriented bonds of the dual lattice, except on those that are dual to the plaquettes tiling a surface bounded by the Polyakov loops, where it takes value $1$ (see also ref.~\cite{Zach:1997yz}, for an analogous calculation in four dimensions). As was shown in ref.~\cite{Caselle:2014eka} (and in refs.~\cite{Panero:2005iu, Panero:2004zq} in the four-dimensional case), the right-hand side of eq.~(\ref{dual_PP_correlator}) can be conveniently factorized in Monte~Carlo calculations, where it can be combined with powerful error-reduction techniques~\cite{Parisi:1983hm, deForcrand:2000fi}: in particular, the ratio of correlators at distances $r+a$ and $r$ can be rewritten as
\begin{equation}
\label{dual_correlator_ratio}
\frac{\langle \Ploop^\star (r+a) \Ploop(0) \rangle}{\langle \Ploop^\star (r) \Ploop(0) \rangle} = \prod_{i=0}^{N_t-1} \frac{\mathcal{Z}_{(N_t \times N_r)+i+1}}{\mathcal{Z}_{(N_t \times N_r)+i}},
\end{equation}
where $\mathcal{Z}_{(N_t \times N_r)+i}$ denotes the partition function of the dual model, in which $n_\nu(y)=1$ only on the $N_t \times N_r$ links dual to the plaquettes between the worldlines of the sources at a distance $r$, and on the first $i$ additional links in the column of plaquettes between $\Ploop^\star (r)$ and $\Ploop^\star (r+a)$. Thus, eq.~(\ref{dual_correlator_ratio}) expresses the ratio of correlators as a product of expectation values of ratios of modified Bessel functions of the first kind of orders differing by one, and argument $\beta$,
\begin{equation}
\label{dual_correlator_ratio_manifestly_local_factors}
H(r)=\frac{\langle \Ploop^\star (r+a) \Ploop(0) \rangle}{\langle \Ploop^\star (r) \Ploop(0) \rangle} = \prod_{i=0}^{N_t-1} \left\langle \frac{I_{|s(x)-s(x+a\hat{\nu})+1|}(\beta)}{I_{|s(x)-s(x+a\hat{\nu})|}(\beta)} \right\rangle_{(N_t \times N_r)+i},
\end{equation}
where the notation $\langle \dots \rangle_{(N_t \times N_r)+i}$ represents an expectation value in the presence of $(N_t \times N_r)+i$ bonds with $n_\nu(y)=1$, and the link from $x$ to $x+a\hat{\nu}$ is dual to the plaquette that is being added, while ``deforming'' the Wilson line at $r$ into the one at $r+a$. Note that each of the factors appearing on the right-hand side of eq.~(\ref{dual_correlator_ratio_manifestly_local_factors}) is manifestly ultralocal, and can be computed to very high numerical precision, even for very large $r$.

To study the profile of the flux tube induced by a pair of static electric sources, we also consider the expectation value of the field strength in the background of two Polyakov lines: the connected correlator of the field-strength component in the direction $\nu$, parallel to the temporal plane through the electric sources
\begin{equation}
\label{el}
\el (x) = \frac{\langle \Ploop^\star (r) \Ploop(0) \El (x) \rangle}{\langle \Ploop^\star (r) \Ploop(0) \rangle} - \langle \El (x) \rangle
\end{equation}
has a very simple expression in the dual formulation of the model:
\begin{equation}
\label{dual_el}
\el (x) = \frac{\langle s(x)-s(x+a\hat{\nu})+ n_\nu(x)\rangle_{N_t \times N_r}}{\sqrt{\beta}}.
\end{equation}
Following an analogous study for the Ising model~\cite{Caselle:1995fh}, in ref.~\cite{Caselle:2016mqu} it was shown that the profile of the flux tube in compact $\U(1)$ gauge theory at zero temperature has an exponential profile: this is what one expects, if the vacuum of the theory is interpreted as a dual superconductor~\cite{Nambu:1974zg, Mandelstam:1974pi, 'tHooft:1979uj}. In addition, we also consider the two-point correlation function
\begin{equation}
\label{el-el_correlator}
Y(x) = \langle \El (x) \El (0) \rangle
\end{equation}
which can be rewritten as
\begin{equation}
\label{dual_el-el_correlator}
Y(x) = \langle [s(x)-s(x+a)] \cdot [s(0)-s(a)] \rangle 
\end{equation}
in the dual formulation.

As mentioned in section~\ref{sec:introduction}, in this work we study the properties of the theory at finite temperature $T$. As is known, a continuous transition takes place at a finite critical temperature $\Tc$ and universality arguments~\cite{Svetitsky:1982gs} suggest that at this critical point, the infrared physics of the system is insensitive to the details of the action of the theory, and becomes equivalent to that of the two-dimensional XY model. This expectation is confirmed by recent lattice calculations~\cite{Borisenko:2013jwa, Borisenko:2015jea}. We extend the numerical investigation of the high-temperature phase of the theory to temperatures above $\Tc$: due to the peculiar features of the XY model, which are reviewed in the following section~\ref{sec:conformal_xy}, universality arguments analogous to those originally discussed in ref.~\cite{Svetitsky:1982gs} allow one to derive exact analytical predictions for various physical quantities in the high-temperature phase of compact $\U(1)$ gauge theory in three dimensions.

\section{The two-dimensional XY model and its conformal-field-theory description}
\label{sec:conformal_xy}

The two-dimensional XY model is a statistical model with many important applications in condensed matter systems, such as Josephson-junction arrays~\cite{Sondhi:1997zz, Beasley:1979po, Resnick:1981kt}, thin layers of superfluid helium~\cite{Nelson:1977zz, Minnhagen:1987zz}, planar ferromagnetic materials~\cite{Sachs:2013ft}, and the roughening transition~\cite{Abraham:1986ss}. It describes two-component real vectors $\vec{s}(x)$, of unit length, defined on the sites $x$ of a square lattice of linear extent $L$ and spacing $a$, and interacting through the Hamiltonian
\begin{equation}
\label{xy_hamiltonian}
H = - J \sum_{\langle x,y \rangle} \vec{s}(x) \cdot \vec{s}(y) = - J \sum_{\langle x,y \rangle} \cos[ \theta(x) - \theta(y)],
\end{equation}
where $\langle x,y \rangle$ denotes nearest-neighbor pairs of sites, $\theta(x)$ is the angle of $\vec{s}(x)$ with respect to an arbitrarily chosen, fixed direction in the two-dimensional real vector space in which the vectors are defined, and the interaction is ferromagnetic when the coupling $J$ is positive. Note that $\theta(x)$ is defined modulo $2\pi$. The model is invariant under a global internal $\Orth(2)$ symmetry, corresponding to rotations of all spins by an arbitrary constant angle.

Let us consider the bidimensional XY model at a temperature $T$, and define the dimensionless parameter $K=J/T$. As is well known, in two dimensions thermal fluctuations always disorder a system with a continuous symmetry~\cite{Mermin:1966fe, Hohenberg:1967zz, Coleman:1973ci}; as a  consequence, the spontaneous magnetization vanishes at all non-zero temperatures:
\begin{equation}
\langle \vec{s} \rangle = 0 \qquad \mbox{for all $T>0$}.
\end{equation}
More detailed information on the behavior of the model in the high-temperature limit can be obtained by a Fourier transform over the internal $\Orth(2)$ group: the calculation shows that at small $K$ the two-point spin correlation function decays exponentially with the spin-spin spatial separation $r$:
\begin{equation}
G(r) = \langle \vec{s}(x) \cdot \vec{s}(y) \rangle \sim \exp \left( -\frac{r}{\xi} \right), \qquad \mbox{with $r=|x-y|$};
\end{equation}
the correlation length $\xi$ is temperature-dependent.

On the other hand, in the low-temperature ($T \to 0$) limit the ferromagnetic nature of the interaction favors spin alignment, thus $\theta$ is expected to be a slowly varying function of space, and the cosine appearing in eq.~(\ref{xy_hamiltonian}) can be approximated by the first two terms in its Taylor expansion:
\begin{equation}
\label{xy_hamiltonian_low_temperature_limit}
H \simeq \frac{J}{2} \sum_{\langle x,y \rangle} \left[ \theta(x) - \theta(y)\right]^2 + \mbox{const}.
\end{equation}
In this limit, the lowest-energy excitations of the system are spin waves: they induce an
\emph{algebraic} decay of $G(r)$ with the spin-spin separation~\cite{McBryan:1977ot, Frohlich:1981yn},
\begin{equation}
\label{cold_xy_correlator}
G(r) \sim \left( \frac{r}{L} \right)^{-\eta},
\end{equation}
where the exponent $\eta$ varies continuously as a function of the temperature, approaching zero linearly in the temperature as $\eta=1/(2\pi K)$ for $T \to 0$~\cite{Berezinsky:1970fr}.

The qualitatively different behavior of $G(r)$ at high and at low temperatures indicates that, while this model does not display genuine long-range order at any finite temperature $T$, it nevertheless admits two different phases: at high temperatures the system is disordered, while at low temperatures it is characterized by a non-conventional ``quasi-long-range'' order, of topological origin. To understand this, we observe that the equation of motion derived from eq.~(\ref{xy_hamiltonian}) admits topologically non-trivial ``vortex'' solutions, in which the $\theta$ field ``winds around'' a given point (the vortex center) an integer number $n$ of times. Vortices satisfy $\oint \nabla \theta \cdot \dd l = 2 \pi n$ for all positively oriented loop encircling the vortex center. The vortex energy goes like $\pi n^2 J \ln (L/a)$, i.e. is proportional to the square of the vortex charge and diverges in the thermodynamic limit $L \to \infty$. By contrast, the energy of a single-charge vortex-antivortex pair separated by a finite distance $r$ remains finite.

Vortices play a key r\^ole in determining the properties of the two phases: as the creation of the core of a vortex requires a finite energy cost $\Ecore$, thermally excited vortices at equilibrium contribute terms proportional to $\exp( -\Ecore/T)$ to the partition function. Moreover, the energy cost of isolated vortices (which is logarithmically divergent with the system size) forces them to remain bound in vortex-antivortex pairs at low temperatures. However, a simple estimate of the single-vortex free energy, neglecting interactions, $F \simeq (\pi J - 2T) \ln (L/a)$, reveals that, as the temperature is increased, the energy cost of an isolated vortex is eventually (over)compensated by entropy, and free vortices start to proliferate at a finite temperature $\TKT$, where an infinite-order transition takes place: the Kosterlitz-Thouless phase transition~\cite{Kosterlitz:1973xp}.

For all temperatures $T>\TKT$ the system behaves as a gas of unbound vortices, interacting with each other through a logarithmic Coulomb potential. The value of the Kosterlitz-Thouless temperature $\TKT$ has been computed numerically to high precision: $\TKT/J=0.89294(8)$~\cite{Hasenbusch:1994hr, Hasenbusch:1996eu, Hasenbusch:2005xm}. In fact, in the low-temperature phase, all effects neglected in the heuristic estimate of the vortex free energy discussed above induce only a \emph{quantitative} correction with respect to the result from the spin-wave approximation.

A more quantitative description of the dynamics of the model can be obtained through the renormalization group, as discussed in detail in the appendix~\ref{app:RG_analysis_of_XY_model}. The main result of this analysis is that, for $\tau=T/\TKT-1 \to 0^+$, the correlation length diverges as
\begin{equation}
\label{xi_singularity}
\frac{\xi}{a} \sim \exp \left(\frac{b}{\sqrt{\tau}}\right),
\end{equation}
with $b$ a non-universal, positive constant, implying that the Kosterlitz-Thouless transition is of infinite order. In addition, one also finds that the large-distance behavior of the two-point correlation function at $T=\TKT$ is of the form
\begin{equation}
\label{critical_correlator}
G(r) \sim \frac{\left( \ln r \right)^{2\Theta}}{r^\eta},
\end{equation}
with $\Theta=1/16$ and $\eta=1/4$~\cite{Pelissetto:2000ek}. 

On the other hand, eq.~(\ref{cold_xy_correlator}) shows that for $T<\TKT$ the two-point correlation function always decays like an inverse power of $r$, with a temperature-dependent exponent~\cite{Berezinsky:1970fr}. Thus, the whole low-temperature phase of the model is characterized by scale-invariant behavior (of Gau{\ss}ian type), and $T=\TKT$ is actually a multicritical point: this can be shown by generalizing the model with two additional parameters, that control the energy cost of introducing a vortex in the model and the coupling to an explicit symmetry-breaking interaction~\cite{Jose:1977gm, Kadanoff:1978pv, Kadanoff:1979mb}.

The scale invariance of the XY model for all temperatures $T<\TKT$ is closely related to the fact that the cold phase of this model admits a conformal-field-theory description in terms of a free, massless compact bosonic field, with central charge $c=1$, which can be identified with the phase $\theta$. The periodicity of this field implies that the theory has both ``electric'' and ``magnetic'' (i.e. ``vortex'') operators, obeying a Dirac quantization condition. It is known that, in a $c=1$ theory, the existence of marginal operators with conformal weights $h=\bar{h}=1$ leads, under appropriate conditions~\cite{Dijkgraaf:1987vp}, to the existence of a continuous line of conformal theories. For the low-temperature phase of the bidimensional XY model, the marginal operator can be associated with the periodicity of the field. In refs.~\cite{Kadanoff:1978pv, Kadanoff:1979mb} it was shown that the operators
\begin{equation}
\label{s_nm}
s_{n,m}=\frac{S_{n,m}+S_{-n,-m}}{\sqrt{2}}
\end{equation}
(where $S_{n,m}$ is an operator creating an excitation with spin-wave index $n$ and vorticity number $m$, for integer $n$ and $m$) have critical indices
\begin{equation}
\label{x_nm}
x_{n,m}=\frac{1}{2} \left( \frac{n^2}{2\pi K} + 2\pi K m^2 \right), \qquad l_{n,m}=-nm
\end{equation}
for the scaling dimension and spin, respectively. Note that there exists an S-duality, interchanging electric and magnetic excitations, which corresponds to exchanging $\sqrt{2\pi K} \to 1/\sqrt{2\pi K}$. The Kosterlitz-Thouless point, at $T=\TKT$, corresponds to $K=2/\pi$, so in the low-temperature phase the scaling dimension of the electric operator $s_{1,0}$ increases continuously with $T$ as $x_{1,0}=1/(4\pi K)$, tending to the critical value $1/8$---see eq.~(\ref{critical_correlator})---for $T \to \TKT^{-}$.

At the Kosterlitz-Thouless temperature, the electric operator $s_{1,0}$ of the XY model in two dimensions can be directly associated with the loop operator $\Ploop$ of the three-dimensional $\U(1)$ gauge theory; similarly, the $s_{2,0}$ (``energy density'') operator has its counterpart in the action density of the gauge theory~\cite{Svetitsky:1982gs}. Finally, the connected correlation function of the field-strength component parallel to the plane through the electric sources defined in eq.~(\ref{el}) is mapped to
\begin{equation}
\label{tube_profile_in_xy}
Q(r,x) = \frac{\langle\vec{s}(r)\vec{s}(0)\phi(x)\rangle}{\langle\vec{s}(r)\vec{s}(0)\rangle} - \langle\phi\rangle,
\end{equation}
where $\phi$ denotes the ``flux'' operator in the XY model: its one-point correlation function vanishes because of scale invariance, hence the last term on the right-hand side of eq.~(\ref{tube_profile_in_xy}) can be dropped. By looking at its symmetries, it is easy to realize that $\phi$ does not correspond to one of the $O_{n,m}$; dimensional analysis and Gau{\ss}' theorem suggest that its scaling dimension is $\Delta_\phi=1$, but this value can be affected by corrections due to interactions. Conformal invariance leads to the prediction
\begin{equation}
\label{tube_profile_cft_prediction}
Q(r,x) = \frac{c_{s s \phi}}{\left(r/4\right)^{\Delta_\phi}} \left( 1+\frac{4x^2}{r^2} \right)^{-\Delta_\phi} = C(r) \left( 1+\frac{4x^2}{r^2} \right)^{-\Delta_\phi},
\end{equation}
where $c_{s s \phi}$ is the coefficient of the $\phi$ term appearing in the operator product expansion of $\vec{s}$ with itself: $\vec{s}(x+\delta x) \cdot \vec{s}(x) \sim c_{s s \phi} \phi(x) + \dots$ for $\delta x \to 0$~\cite{Wilson:1969zs}.\footnote{Alternatively, one can also define the quantity denoted as $c^{\mbox{\tiny{cont}}}_{s s \phi}$, which plays the same r\^ole as $c_{s s \phi}$ but for the continuum counterpart of $\phi$, assuming the amplitude of its two-point correlation function to be fixed to $1$.} For later convenience, we also define $C(r)=c_{s s \phi} \cdot \left(r/4\right)^{-\Delta_\phi}$.

As the mapping between operators in the three-dimensional $\U(1)$ gauge theory and those in the two-dimensional XY model is based only on the existence of an infinite correlation length (not on the presence of a phase transition), we tested whether it can be extended to the whole high-temperature phase of the $\U(1)$ gauge theory, which is expected to correspond to the low-temperature phase of the XY model. More precisely, we compared the results of a set of high-precision lattice calculations for the $\U(1)$ theory with the analytical conformal-field-theory predictions for the XY model. Our results are presented in the following section~\ref{sec:results}.

\section{Numerical results}
\label{sec:results}

Using the dual formulation of the theory, we carried out Monte~Carlo calculations of compact $\U(1)$ gauge theory on isotropic lattices of spacing $a$ and volume $(N_s^2 \times N_t)a^3$. In order to limit the impact of finite-volume effects, all simulations were carried out in the $N_s \gg N_t$ regime (with $N_s/N_t$ typically larger than $30$). Specifically, for $N_t=4$ the value of $\beta=1/(ae^2)$ corresponding to the critical temperature is $\beta=3.005$~\cite{Borisenko:2015jea}, and $\beta=4$ is significantly higher than the critical temperature. The setup of our simulations is summarized in table~\ref{tab:sim_setup}, where $r$ denotes the distance between the Wilson lines $\Ploop$, that wind around the Euclidean-time direction, are oppositely oriented, and separated along one of the main spatial axes of the lattice. $x$, on the other hand, denotes the distance from the plane of the Wilson lines, at which we probe the flux-tube profile, by calculating the expectation value of the electric-field component defined in eq.~(\ref{el}).

\begin{table}
\centering
\begin{tabular}{|c|c|c|c|c|c|c|}
\hline
ensemble & $N_s$ & $N_t$ & $\beta=1/(ae^2)$ & statistics        & $r/a$               & $x/a$     \\
\hline
A        & $192$ & $4$   & $3.010$          & $6.4 \times 10^5$ & $[0,31]$            &    --     \\
         & $192$ & $4$   & $3.025$          & $6.4 \times 10^5$ & $[0,31]$            &    --     \\
         & $192$ & $4$   & $3.050$          & $6.4 \times 10^5$ & $[0,31]$            &    --     \\
         & $192$ & $4$   & $3.080$          & $6.4 \times 10^5$ & $[0,31]$            &    --     \\
         & $192$ & $4$   & $3.125$          & $6.4 \times 10^5$ & $[0,31]$            &    --     \\
         & $192$ & $4$   & $3.250$          & $6.4 \times 10^5$ & $[0,31]$            &    --     \\
         & $192$ & $4$   & $3.500$          & $6.4 \times 10^5$ & $[0,31]$            &    --     \\
         & $192$ & $4$   & $3.750$          & $6.4 \times 10^5$ & $[0,31]$            &    --     \\
         & $192$ & $4$   & $4.000$          & $6.4 \times 10^5$ & $[0,31]$            &    --     \\
         & $192$ & $4$   & $4.100$          & $6.4 \times 10^5$ & $[0,31]$            &    --     \\
\hline
B        & $128$ & $4$   & $4.0$            & $6.4 \times 10^5$ & $[10,80]$           & $[2,20]$  \\
         & $192$ & $4$   & $4.0$            & $6.4 \times 10^5$ & $[10,80]$           & $[2,30]$  \\
         & $256$ & $4$   & $4.0$            & $8.0 \times 10^6$ & $[10,80]$           & $[2,40]$ \\
         & $320$ & $4$   & $4.0$            & $3.2 \times 10^6$ & $60,80$             & $[2,40]$ \\
\hline
\end{tabular}
\caption{Setup of our simulations.}
\label{tab:sim_setup}
\end{table}

The main goal of our analysis consists in testing whether the conjecture that the long-wavelength properties of the $\U(1)$ gauge theory at temperatures $T \ge \Tc$ can be described by the low-temperature phase of the XY model in two dimensions holds or not.

\begin{figure}[h!]
\begin{center}
\includegraphics*[width=\textwidth]{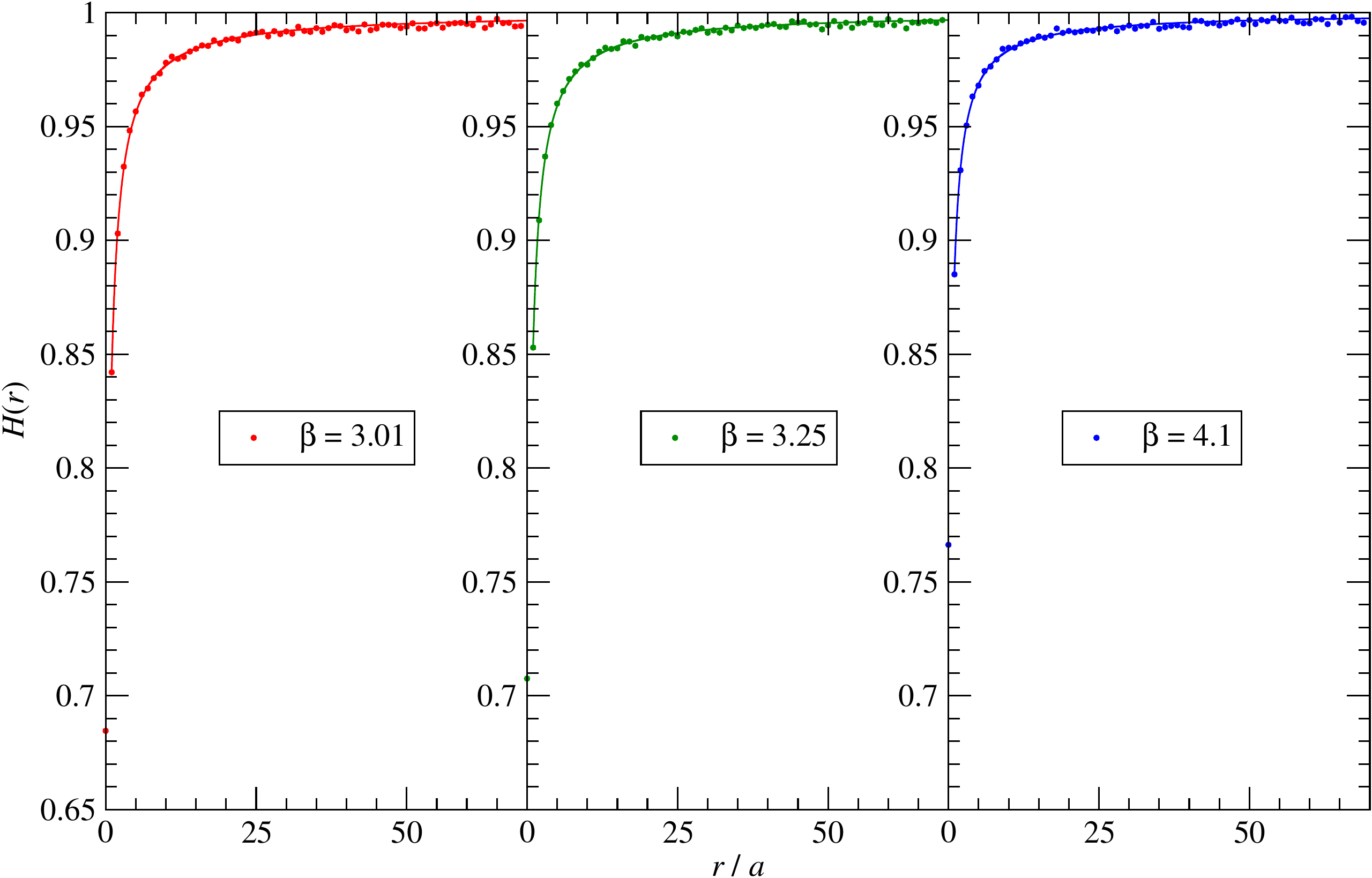}
\caption{\label{fig:correlator_ratios} Example of results obtained for correlator ratios, from simulations at three different values of $\beta=1/(ae^2)$. The data from numerical lattice calculations of $\U(1)$ gauge theory in three dimensions are compared with one-parameter fits to the functional form in eq.~(\ref{correlator-ratio_prediction}), derived from the conformal-field-theory description of the XY model in two dimensions.}
\end{center}
\end{figure}

\begin{figure}[h!]
\begin{center}
\includegraphics*[width=\textwidth]{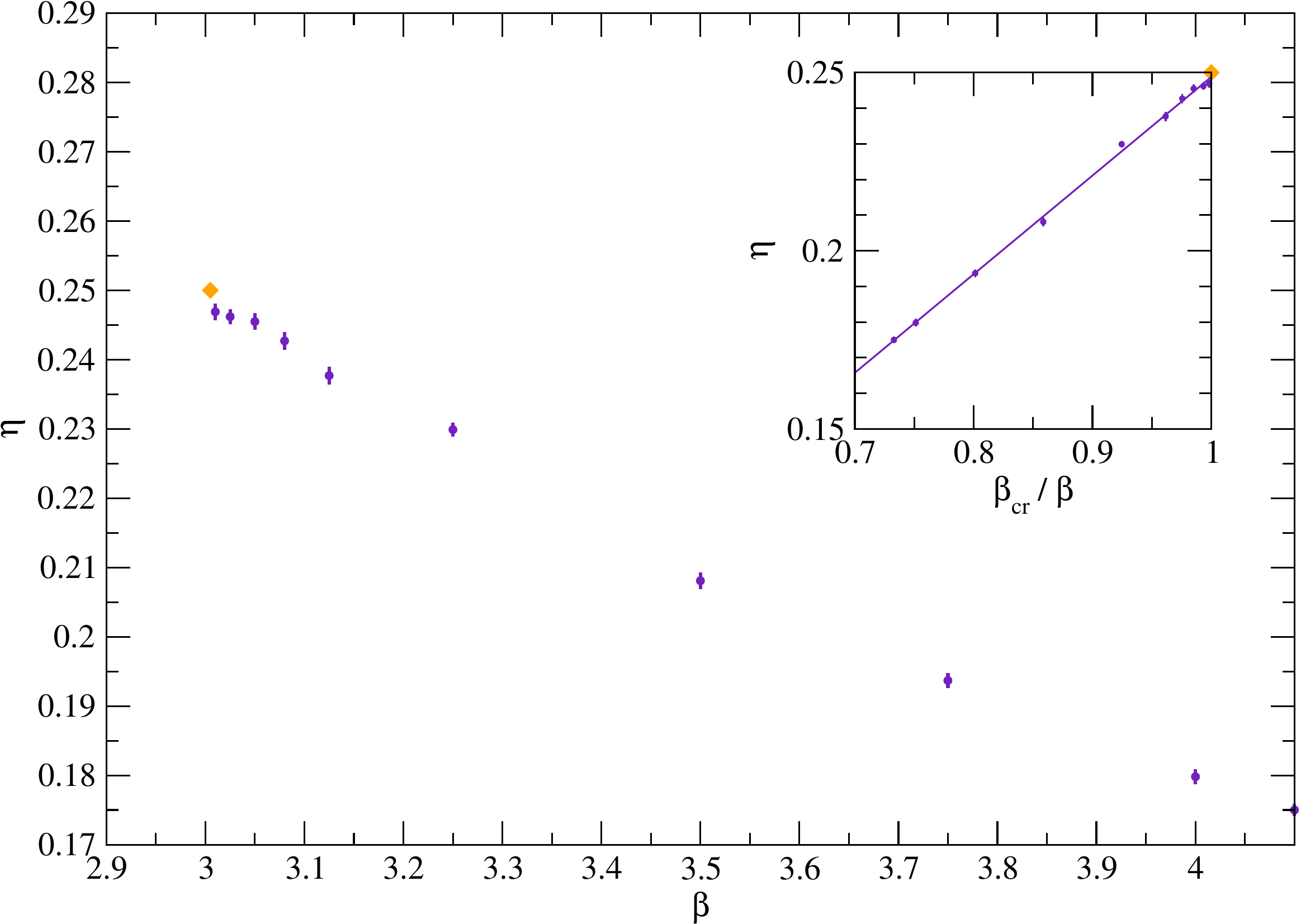}
\caption{\label{fig:eta_versus_beta} Results obtained for $\eta$, in simulations at different values of $\beta$ and for $N_s=192$ and $N_t=4$, from fits of Wilson-line-correlator ratios to eq.~(\ref{correlator-ratio_prediction}). The orange diamond indicates the conformal-field-theory prediction $\eta=1/4$ at the critical temperature, which corresponds to $\beta=3.005$~\cite{Borisenko:2015jea}. In the inset, the same data are shown as a function of $\betac/\beta$ and fitted to eq.~(\ref{eta_betac_over_beta}), with the rationale discussed in the text.}
\end{center}
\end{figure}

To this purpose, we first studied the behavior of the $\langle \Ploop^\star (r) \Ploop(0) \rangle$ correlator, whose long-distance behavior is expected to be described by eq.~(\ref{cold_xy_correlator}). We computed ratios of Wilson-line correlators using eq.~(\ref{dual_correlator_ratio_manifestly_local_factors}) and compared them to the prediction
\begin{equation}
\label{correlator-ratio_prediction}
H(r)=\frac{\langle \Ploop^\star (r+a) \Ploop(0) \rangle}{\langle \Ploop^\star (r) \Ploop(0) \rangle} = \left( 1 + \frac{a}{r} \right)^{-\eta}
\end{equation}
by means of one-parameter fits, with $\eta$ as the fitted parameter. Three examples of these fits are shown in fig.~\ref{fig:correlator_ratios}, while the complete results are reported in table~\ref{tab:eta_versus_beta} and displayed in fig.~\ref{fig:eta_versus_beta}. Remarkably, even though in our fits we discarded the data at small values of $r$ (which are expected to be affected by lattice-discretization effects\footnote{At the quantitative level, we observe that the fit results, in particular those obtained from data sets at the $\beta$ values closest to $\betac$, exhibit some dependence on the smallest value of $r$ that is included in the fit: the induced systematic uncertainty on $\eta$ is at most of the order of a few per mille.}), the curves obtained from these fits follow closely our numerical results down to values of $r/a=O(1)$.

The analysis shows that, in all cases, the ratios of Wilson-line correlators can be perfectly fitted to the expected functional form, indicating very clearly that in the high-temperature phase of $\U(1)$ gauge theory $\langle  \Ploop^\star (r) \Ploop(0) \rangle$ decays as a power of $r$. All values of the reduced $\chi^2$ (listed in the fifth column) for these one-parameter fits are close to $1$, and the statistical uncertainty on the fitted parameter $\eta$ is of the order of five per mille.

The analysis also shows that, when $\beta$ is close to the critical value, $\eta$ tends to $1/4$, as predicted by conformal field theory.

Another interesting observation from our analysis concerns the relation between the parameters of the $\U(1)$ gauge theory and those of the XY model describing the same long-wavelength physics. On very general grounds, it is known that the $T\to\Tc^+$ limit of a gauge theory with a continuous thermal transition corresponds to the $T\to\Tc^-$ limit of the spin model~\cite{Svetitsky:1982gs}. For the $\U(1)$ gauge theory in three dimensions, in which the whole high-temperature phase is mapped to the low-temperature phase of the XY model, one then expects the long-distance physics at higher and higher temperatures (i.e. further and further away from $\Tc$) to be captured by the XY model at lower and lower temperatures (that is, further and further away from $\TKT$). Considering the dimensionless ratios $T/\Tc$ in the three-dimensional $\U(1)$ gauge theory and $T/\TKT$ in the bidimensional XY model, it is thus tempting to think that the theories describe the same infrared physics when these two ratios are (approximately) the inverse of each other.

To test this hypothesis at a quantitative level, we make two further observations. Firstly, the temperature of the $\U(1)$ lattice gauge theory is given by $T=1/(a N_t)$. Since the squared coupling in a three-dimensional gauge theory has dimension one, the inverse lattice spacing $1/a$ (and, as a consequence, $T$) is approximately proportional to $\beta=1/(ae^2)$; this relation is not expected to be exact, due to quantum corrections. Secondly, the properties of the bidimensional XY model at finite temperature can be conveniently described by introducing the spin-wave stiffness $\rhos$ (see the appendix~\ref{app:RG_analysis_of_XY_model}): then, the physics of the model is determined by the dimensionless parameter $K=\rhos/T$, which reduces to $J/T$ for $T \to 0$, and accounts for effects related to the density of vortices at finite temperature. In particular, in the low-temperature phase, the $G(r)$ correlator decays as described by eq.~(\ref{cold_xy_correlator}), with $\eta=1/(2\pi K)$. This means that (neglecting the fact that $\rhos$ is not a constant, but rather a temperature-dependent quantity) $\eta$ is expected to be approximately proportional to the temperature of the XY model. Combining these pieces of information (with the approximations mentioned), one can thus expect the exponent $\eta$ extracted from the numerical results for $\langle \Ploop (r) \Ploop (0) \rangle$ correlators in the $\U(1)$ lattice gauge theory at $1<T/\Tc \simeq \beta/\betac$ to be equal to the exponent $\eta$ of the XY model at $T/\TKT=\betac/\beta$, i.e. to have a linear dependence between $\eta$ and $\betac/\beta$. Fitting our results to
\begin{equation}
\label{eta_betac_over_beta}
\eta = a_1 \frac{\betac}{\beta} +a_0,
\end{equation}
we obtain $a_1=0.2767(35)$ and $a_0=-0.0279(31)$, with $8$ degrees of freedom and $\redchisq=1.04$, indicating excellent agreement with this crude model. The result of this analysis is shown in the inset plot in figure~\ref{fig:eta_versus_beta}.

\begin{table}
\centering
\begin{tabular}{|c|c|c|c|c|c|}
\hline
$N_s$ & $N_t$ & $\beta$ & $\eta$       & $\redchisq$ & d.o.f. \\
\hline
$192$ & $4$   & $3.010$ & $0.2469(12)$ & $1.19$      & $30$   \\
$192$ & $4$   & $3.025$ & $0.2462(11)$ & $1.02$      & $30$   \\
$192$ & $4$   & $3.050$ & $0.2455(12)$ & $1.18$      & $30$   \\
$192$ & $4$   & $3.080$ & $0.2427(13)$ & $1.36$      & $30$   \\
$192$ & $4$   & $3.125$ & $0.2377(13)$ & $1.59$      & $30$   \\
$192$ & $4$   & $3.250$ & $0.2299(10)$ & $0.86$      & $30$   \\
$192$ & $4$   & $3.500$ & $0.2081(12)$ & $1.44$      & $30$   \\
$192$ & $4$   & $3.750$ & $0.1937(11)$ & $1.39$      & $30$   \\
$192$ & $4$   & $4.000$ & $0.1798(11)$ & $1.45$      & $30$   \\
$192$ & $4$   & $4.100$ & $0.1750(9)$  & $0.98$      & $30$   \\
\hline
\end{tabular}
\caption{Fits of $H(r)$ to eq.~(\ref{correlator-ratio_prediction}), from simulations on a lattice of volume $(N_s^2 \times N_t) a^3$, with $N_s=192$ and $N_t=4$, at different values of $\beta=1/(ae^2)$.}
\label{tab:eta_versus_beta}
\end{table}

\begin{figure}[h!]
\begin{center}
\includegraphics*[width=\textwidth]{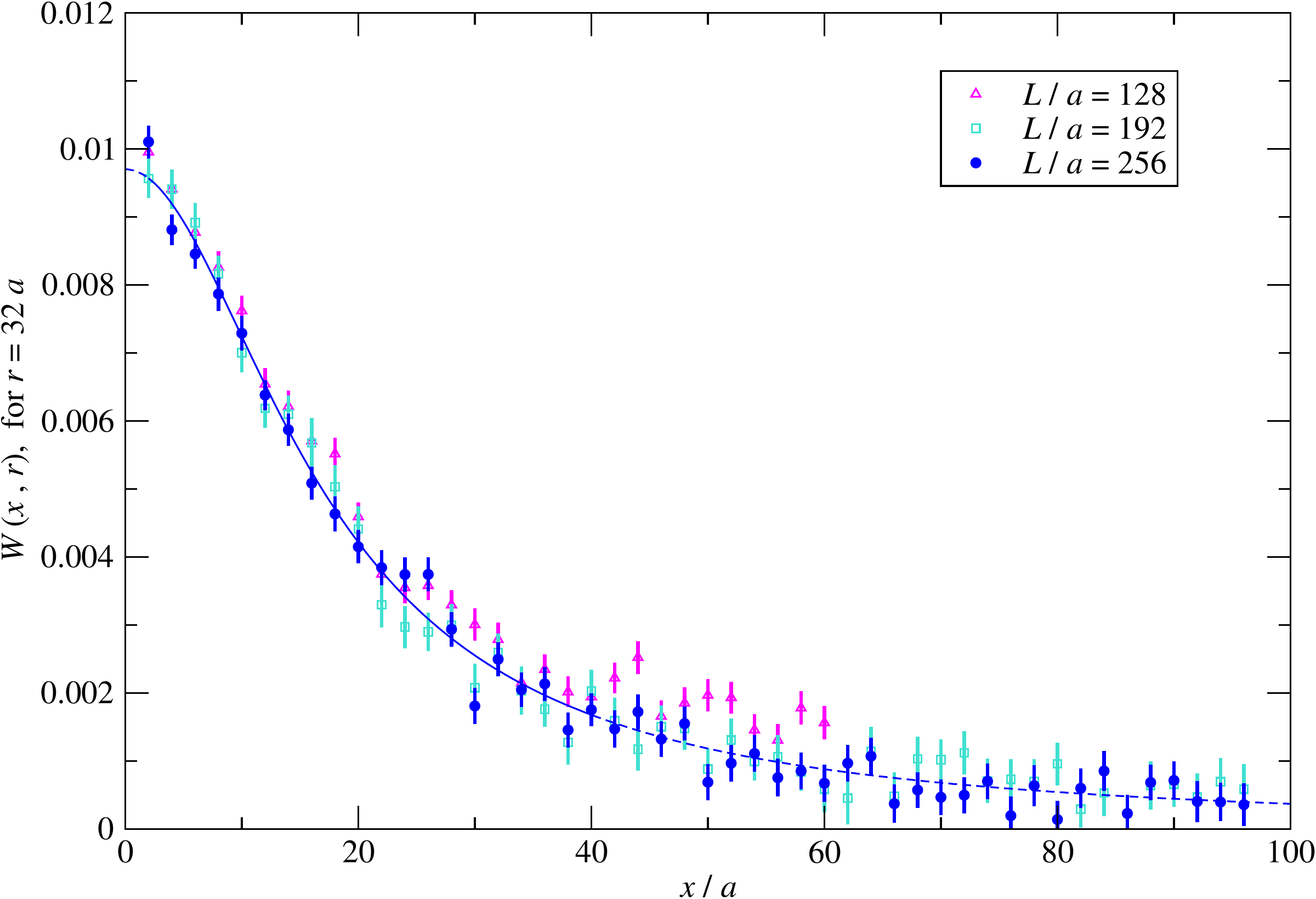}
\caption{\label{fig:flux_tube} Flux-tube profile, as probed by the quantity defined in eq.~(\ref{el}), from calculations at $\beta=4$, with $N_t=4$ and $r=32a$, for $N_s=128$ (magenta triangles), $N_s=192$ (turquoise squares), and $N_s=256$ (blue circles). The blue curve is the fit of the $N_s=256$ data to eq.~(\ref{tube_profile_cft_prediction}) in the $2 \le x/a \le 40$ range.}
\end{center}
\end{figure}

Next, we analyzed the flux-tube profile, probed by the operator defined in eq.~(\ref{el}), and compared our numerical results with the conformal-field-theory prediction given by eq.~(\ref{tube_profile_cft_prediction}). Fig.~\ref{fig:flux_tube} shows an example of this analysis, focusing on our simulations at $\beta=4$, $N_t=4$ (corresponding to a temperature significantly higher than $\Tc$): the numerical results for $W(x,r)$, at fixed $r=32a$, are plotted as a function of the distance (in lattice-spacing units) from the plane through the Wilson lines. The figure shows that the results for three different values of the spatial linear extent of the lattice ($L=128a$, $L=192a$, and $L=256a$) are essentially compatible with each other: this indicates that this quantity is not strongly affected by finite-size effects.\footnote{Note that, in contrast to the $G(r)$ correlator, finite-size corrections and the effects of periodic images of the lattice would be difficult to account for properly, since they have a different impact on the source worldlines and on the flux operator. To this purpose, we chose to restrict our analysis to values $L/x \ge 5$.} The data for $L=256a$ in the range $2 \le x/a \le 40$ are fitted to eq.~(\ref{tube_profile_cft_prediction}), using $C(r)$ and $\Delta_\phi$ as fit parameters, and the result of this fit is the blue curve: the fitted parameters are $C(r)=0.0097(1)$ and $\Delta_\phi=0.886(22)$, with $\redchisq=1.76$. It is remarkable that the agreement of the fitted curve extends well beyond the interval of fitted data (solid line), both down to shorter and up to larger values of $x/a$ (dashed portions of the curve).

Table~\ref{tab:Delta_phi_from_flux} shows a more complete summary of these two-parameter fits. As one can see, when $r$ is increased, $C(r)$ and the precision on $\Delta_\phi$ decrease. Nevertheless, the precision of the results is sufficient to observe that essentially all results for $\Delta_\phi$, reported in the fifth column, are compatible with each other: a fit to a constant yields $\Delta_\phi=0.933(7)$, with $\redchisq=1.87$, the only outliers being the results obtained for $r=32a$ and $r=80a$.

\begin{table}
\centering
\begin{tabular}{|c|c|c|c|c|c|}
\hline
 $N_s$ & $N_t$ & $r/a$ & $C(r)$      & $\Delta_\phi$ & $\redchisq$ \\
\hline
 $256$ & $4$   & $10$  & $0.0297(2)$ & $0.925(11)$   & $ 0.86$     \\
 $256$ & $4$   & $16$  & $0.0196(2)$ & $0.959(14)$   & $ 1.57$     \\
 $256$ & $4$   & $20$  & $0.0154(2)$ & $0.944(17)$   & $ 0.77$     \\
 $256$ & $4$   & $32$  & $0.0097(1)$ & $0.886(22)$   & $ 1.76$     \\
 $256$ & $4$   & $40$  & $0.0078(1)$ & $0.935(31)$   & $ 0.81$     \\
 $256$ & $4$   & $50$  & $0.0063(1)$ & $0.926(39)$   & $ 0.58$     \\
 $320$ & $4$   & $60$  & $0.0056(2)$ & $1.03(9)$     & $ 1.79$     \\
 $320$ & $4$   & $80$  & $0.0042(1)$ & $0.79(8)$     & $ 0.51$     \\
\hline
\end{tabular}
\caption{Results of the fits of eq.~(\ref{tube_profile_cft_prediction}), with $C$ and $\Delta_\phi$ as fit parameters.
}
\label{tab:Delta_phi_from_flux}
\end{table}

\begin{figure}[h!]
\begin{center}
\includegraphics*[width=\textwidth]{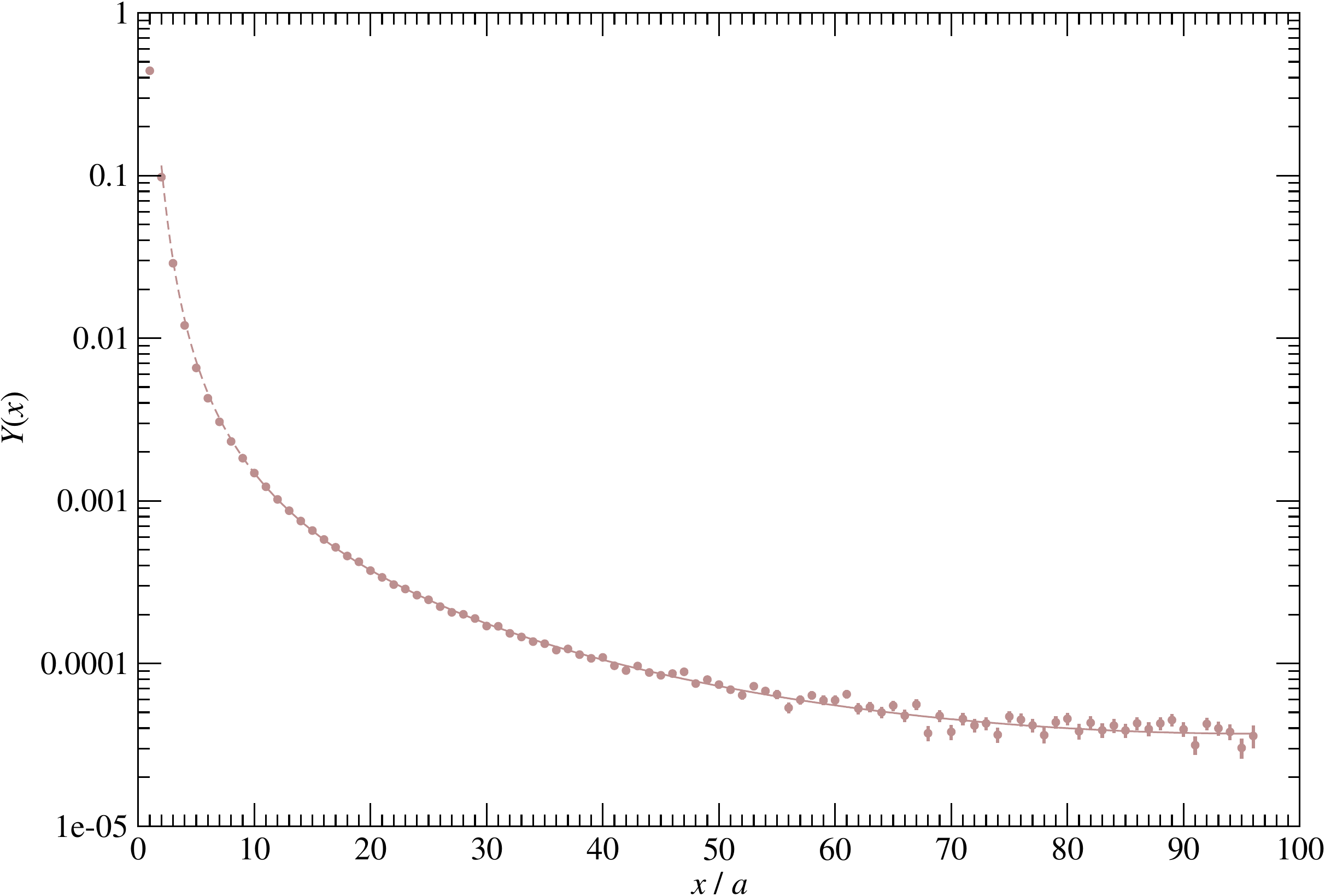}
\caption{\label{fig:el-el_correlator} Field-strength two-point correlation function, computed according to eq.~(\ref{dual_el-el_correlator}), for $\beta=4$, $N_s=192$, and $N_t=4$. The solid curve is obtained from the fit to eq.~(\ref{Y_fit}), as discussed in the text.}
\end{center}
\end{figure}

A different, and independent, way to extract $\Delta_\phi$ is based on the analysis of the $Y(x)$ correlator, evaluated according to eq.~(\ref{dual_el-el_correlator}): see figure~\ref{fig:el-el_correlator} for an example of numerical results, for $\beta=4.0$, $N_s=192$, and $N_t=4$. As the lattice realization of the field strength generally involves mixing of different operators, we fit our results for $Y(x)$ to the functional form
\begin{equation}
\label{Y_fit}
Y(x) = b_0 \cdot \left[ \left( \frac{a}{x} \right)^{2 \Delta_\phi} + \left( \frac{a}{L-x}\right)^{2 \Delta_\phi}\right] + b_1 \cdot \left[ \left(\frac{a}{x}\right)^4 + \left(\frac{a}{L-x}\right)^4 \right],
\end{equation}
which also includes the leading corrections due to the finite spatial extent of the system and the contribution of the operator with conformal weight $2$ (associated with the action density). It is important to stress that the conformal weight of this marginal operator is exact~\cite{Ginsparg:1988ui}. The three-parameter fit of these data in the $10 \le x/a \le 96$ range yields $b_0=0.1046(33)$, $b_1=1.40(16)$, and $\Delta_\phi=0.946(5)$, with $\redchisq=1.15$. It is interesting to note that $b_1/b_0=O(10)$: the large value of the $b_1$ coefficient implies that, while the behavior of the correlation function at large distances is dominated by the flux operator, with conformal weight $\Delta_\phi \simeq 1$, the mixing with the operator of conformal weight $2$ induces a non-negligible correction at short and intermediate distances. 

We remark that the result for $\Delta_\phi$ from this analysis is essentially consistent with the one obtained from the study of $W(x,r)$, which is based on a different type of operator, evaluated on a different set of configurations.

We also observe that, if the results for $W(x,r)$ (for $\beta=4$ and $N_t=4$) are fitted to eq.~(\ref{tube_profile_cft_prediction}) at fixed $\Delta_\phi=0.946(5)$, with $C(r)$ as the only fit parameter, one obtains the results listed in table~\ref{tab:C_from_flux}, in which the statistical uncertainty from the fit is reported in the first parentheses, while the error induced by the uncertainty on $\Delta_\phi$ is given in the second parentheses. The fitting range is $2 \le x/a \le 24$ for all values of $r/a$. Once again, the fits yield good $\chi^2$ values, even including data at small $x$. Note that in this case the possible contamination with the operator of scaling dimension $2$ is expected to be completely negligible, being suppressed as $1/r^2$; this is indeed confirmed: we verified that if such term is included in the fit, its contribution is always compatible with zero, within the precision of our data.

\begin{table}
\centering
\begin{tabular}{|c|c|c|c|}
\hline
 $N_s$  & $r/a$ & $C(r)$            & $\redchisq$ \\
\hline
 $256$  & $10$  & $0.03002(20)(29)$   & $ 0.99$   \\
 $256$  & $16$  & $0.01944(16)(17)$   & $ 1.54$   \\
 $256$  & $20$  & $0.01541(12)(12)$   & $ 0.73$   \\
 $256$  & $32$  & $0.00990(10)(5)$    & $ 2.01$   \\
 $256$  & $40$  & $0.00785(8)(3)$     & $ 0.78$   \\
 $256$  & $50$  & $0.00634(8)(2)$     & $ 0.56$   \\
 $320$  & $60$  & $0.00552(11)(2)$    & $ 1.74$   \\
 $320$  & $80$  & $0.00437(8)(1)$     & $ 0.64$   \\
\hline
\end{tabular}
\caption{One-parameter fits of $W(x,r)$ (for $\beta=4$ and $N_t=4$) to eq.~(\ref{tube_profile_cft_prediction}), with $\Delta_\phi=0.946(5)$ fixed.}
\label{tab:C_from_flux}
\end{table}

Yet another test of the conformal-field-theory prediction for the shape of the flux tube concerns the dependence of $C$ on $r$: fitting the values of $C(r)$ in table~\ref{tab:C_from_flux} to the expected form
\begin{equation}
\label{C_as_a_function_of_r}
C(r) = c_{s s \phi} \cdot \left(\frac{4a}{r}\right)^{\Delta_\phi},
\end{equation}
one obtains $c_{s s \phi}=0.07115(7)$ and $\Delta_\phi=0.948(6)$ with $\redchisq=1.98$, when all data for $10 \le r/a \le 80$ are included in the fit. Finally, the value of $c_{s s \phi}$ with the alternative (``continuum'') normalization of the field operator mentioned in section~\ref{sec:conformal_xy} reads $c^{\mbox{\tiny{cont}}}_{s s \phi}=0.2200(34)$.

\section{Discussion and concluding remarks}
\label{sec:conclusions}

In this work, we presented a high-precision study of $\U(1)$ gauge theory in three spacetime dimensions, in its compact formulation on a cubic lattice---a theory that, as we discussed in section~\ref{sec:introduction}, has important implications for high-energy elementary-particle physics and for condensed-matter physics alike.

As is known, a semiclassical analysis shows that at zero and at low temperatures, the monopoles (or instantons) of the theory are responsible for the dynamical generation of a finite mass gap and induce a logarithmically confining potential for pairs of probe charges~\cite{Polyakov:1976fu, Gopfert:1981er}. These properties persist up to a finite critical temperature $\Tc$, at which the theory undergoes a transition to a different phase, characterized by restoration of scale invariance (at least for modes of wavelength much longer than the lattice spacing) and by logarithmic confinement.

We focused on the dynamics of the theory in this high-temperature regime, and compared a new set of numerical results from Monte~Carlo calculations on high-performance computing machines with analytical predictions from conformal field theory: specifically, we exploited ideas related to universality and to the construction of a low-energy effective field theory for systems characterized by at least one diverging correlation length~\cite{Svetitsky:1982gs} to map \emph{the whole} high-temperature phase of this theory to the the low-temperature phase of the XY model in two dimensions. While the latter is not an exactly solvable model, its properties have been studied for many years and are well understood: in particular, a suitable generalization of the model reveals that the Kosterlitz-Thouless critical point lies at the intersection of different critical lines, with continuously varying critical indices~\cite{Kadanoff:1979mb}. One of these is the line of Gau{\ss}ian critical indices described by eq.~(\ref{x_nm}): their simplicity and the fact that they depend on just one real parameter (e.g. the temperature of the system) allow one to formulate stringent predictions for correlation functions in the XY model. As we showed here, such predictions entail remarkable implications for the correlation functions of the $\U(1)$ theory in its high-temperature phase. While previous work mainly focused on the properties of the theory at or very close to the critical temperature~\cite{Borisenko:2015jea}, here, for the first time, we extended this theoretical machinery to temperatures well above $\Tc$.

The physical quantities that we considered in detail are the two-point correlation functions of probe-source worldlines, and the flux they induce. Using the traditional dual formulation of this gauge theory, a toolbox of powerful algorithmic techniques, and high-performance-computing machines, we were able to track the behavior of these correlators over very long distances, and to obtain very high, and nearly constant, levels of numerical precision for quantities varying over several orders of magnitude. One such example is provided by our analysis of the field-strength two-point correlation function shown in fig.~\ref{fig:el-el_correlator}.

The results that we obtained fully confirm the analytical predictions from conformal field theory, and the validity of the mapping from operators in the high-temperature phase of the gauge theory to operators in the low-temperature phase of the XY model. For $T>\Tc$, the Polyakov-loop correlator in the $\U(1)$ gauge theory decays as an inverse power of the distance, with a characteristic exponent $\eta$ that varies continuously with the temperature. Further analysis, based on a semi-heuristic argument, also suggests that, at least in the temperature range considered here, the corresponding temperatures in the gauge theory and in the bidimensional XY model are approximately inversely proportional to each other. Similarly, a thorough study of the flux tube induced by a pair of static probe charges confirms that its dependence on the spatial charge-charge separation $r$ and on the distance $x$ from the charge-charge axis is accurately described by the functional form predicted by conformal field theory. The critical index $\Delta_\phi$ associated with the ``flux'' operator, which cannot be directly identified with any of the operators defined in eq.~(\ref{s_nm}), is found to differ from its ``classical'' value $1$ by a small but finite negative correction $O(10^{-1})$, i.e. to be renormalized as a result of quantum interactions. This result is confirmed by the analysis of the two-point flux correlation function, in which we detected the leading correction due to the operator of conformal weight $2$ (which can be associated with the cosine of the phase of the plaquette, appearing in Wilson's action): the two different lattice operators mix with each other, and the large coefficient of the operator of conformal weight $2$ makes its contribution non-negligible at short and intermediate distances. As a by-product of our analysis, we also extracted the value of the $c_{s s \phi}$ coefficient, that appears in one of the operator-product expansions relevant for the model.

Our results provide a non-trivial test of the conjecture first put forward in ref.~\cite{Svetitsky:1982gs}, that here, for the first time, is successfully checked in a whole phase of a gauge theory.

We envisage many different directions, in which the approach developed in this paper could also be applied. 

High-precision numerical calculations, like the ones presented here, could also be carried out to study the behavior of this model in the presence of a constant and uniform background electric field, which is expected to have interesting implications for the electric Mei{\ss}ner effect and for the dielectric permittivity of superinsulators~\cite{Diamantini:2018mjg}. Enforcing a smooth background field strength on a periodic lattice in a gauge-invariant way implies a quantization condition on the values of the field that can be studied, but the techniques to perform this type of calculations are well understood and in recent years have been extensively applied to study the effect of external QED fields on strongly interacting matter at high temperature~\cite{DElia:2012ems, Endrodi:2014vza}.

Another interesting generalization is the inclusion of fermionic matter fields. As we already mentioned above, when the theory is coupled to a sufficiently large number $\nf$ of dynamical charged fermion species, its long-distance properties can be described by a strongly coupled conformal field theory: this was shown to be the case in the large-$\nf$ limit~\cite{Appelquist:1988sr}, and is expected to persist also at finite values of $\nf$. It would be very interesting to perform Monte~Carlo calculations of three-dimensional $\U(1)$ gauge theory coupled to dynamical fermions and to perform a systematic comparison of the numerical results with analytical predictions from conformal field theory.

It is worth remarking that inclusion of fermions in this theory comes with some subtleties. In particular, in a three-dimensional (or, more generally, odd-dimensional) spacetime, the parity transformation $\sf{P}$, defined as inversion of all spatial coordinates, is in the group of spatial rotations, hence one defines a different discrete symmetry $\sf{R}$, which inverts only one spatial coordinate. Classically, $\U(1)$ gauge theory defined in three spacetime dimensions coupled to one species of massless Dirac fermions of charge $1$ (in units of $e$) is invariant under $\sf{R}$, but this symmetry is anomalous, i.e. the theory cannot be quantized in a gauge-invariant, $\sf{R}$-preserving way~\cite{Niemi:1983rq, Redlich:1983dv, AlvarezGaume:1984nf}. This anomaly, however, is absent when $\nf$ is even---or a multiple of $4$, on non-orientable manifolds~\cite{Witten:2016cio}.

Adding interacting fermions to this model is also interesting for another reason, namely the rich network of dualities that arise in quantum field theory in $2+1$ dimensions~\cite{Seiberg:2016gmd, Hsin:2016blu, Murugan:2016zal, Karch:2016sxi, Karch:2016aux, Kachru:2016rui, Kachru:2016aon, Aharony:2016jvv, Benini:2017dus, Wang:2017txt, Gaiotto:2017tne, DiPietro:2019hqe}. Such dualities can be considered as a generalization of the conventional particle/vortex duality~\cite{Peskin:1977kp, Dasgupta:1981zz}, and are reviewed in ref.~\cite{Senthil:2018cru}; an analogous web of dualities arises in two dimensions~\cite{Karch:2019lnn}. In particular, it is known that a free electronic Dirac cone is dual to quantum electrodynamics in three dimensions with a single species of fermions and with a ``mixed'' Chern-Simons term that couples a background Abelian gauge field and a dynamical one~\cite{Son:2015xqa, Mross:2015idy}. This duality has important applications in condensed-matter theory, in particular for metallic surfaces of topological insulators and for Fermi liquids induced by strong magnetic fields at half filling of the lowest Landau level~\cite{Wang:2015qmt, Metlitski:2015eka}.

Another interesting topic to be studied is the behavior of equilibrium thermodynamic quantities in three-dimensional compact $\U(1)$ lattice gauge theory. On the one hand, its $T<\Tc$ phase shares many qualitative features with non-Abelian gauge theories in $3+1$~\cite{Boyd:1996bx, Borsanyi:2012ve, Caselle:2018kap, Panero:2009tv, Bruno:2014rxa, Caselle:2015tza, Giusti:2016iqr, Kitazawa:2016dsl, Giudice:2017dor, Iritani:2018idk} or in $2+1$~dimensions~\cite{Christensen:1991rx, Bialas:2008rk, Caselle:2011fy, Caselle:2011mn}. On the other hand, the properties of the theory at $T>\Tc$ are very different from those of its non-Abelian counterpart (most remarkably, the high-temperature phase is confining).

Finally, it would also be interesting to repeat the present study at larger values of $N_t$, and to study quantitatively the dependence of the results on this parameter. Addressing this  issue, however, is clearly beyond the scope of this article, and we leave it for future work.

\vskip1.0cm 
\noindent{\bf Acknowledgements}\\
We thank Maria~Cristina~Diamantini and Carlo~Trugenberger for helpful discussions. The simulations were run on the MARCONI supercomputer of the Consorzio Interuniversitario per il Calcolo Automatico dell'Italia Nord Orientale (CINECA). D.~V. acknowledges support from the INFN~HPC{\_}HTC project.

\begin{appendix}
\section{Renormalization-group analysis of the XY model in two dimensions}
\label{app:RG_analysis_of_XY_model}
\renewcommand{\theequation}{A.\arabic{equation}}
\setcounter{equation}{0}

In this appendix, we discuss the analysis of the XY model in two dimensions by the renormalization group and present a derivation of eq.~(\ref{xi_singularity}).

For a renormalization-group analysis of the XY model in two dimensions, it is convenient to consider the variation in free-energy density $f=F/L^2$ that is induced by imposing a gradient $v$ to the phase field $\theta(x)$:
\begin{equation}
\theta(x) \to \theta(x) + v \cdot x,
\end{equation}
and to introduce a quantity $\rhos$, called the spin-wave stiffness, defined as the second derivative of $f$ with respect to $v$. From eq.~(\ref{xy_hamiltonian}) it follows that at $T=0$ the spin-wave stiffness equals $J$. At finite temperatures below $\TKT$, $\rhos$ can be expressed in terms of the zero-momentum limit of the Fourier transform $\tilde{n}(q)$ of the vortex density:
\begin{equation}
\label{renormalized_spin-wave_stiffness}
\rhos = J - \frac{(2\pi J)^2}{T} \lim_{q \to 0} \frac{\langle \tilde{n}(q) \tilde{n}(-q) \rangle}{q^2}.
\end{equation}
Introducing the dimensionless ratio $K_0=J/T$ and its counterpart $K=\rhos/T$, which accounts for thermal effects, eq.~(\ref{renormalized_spin-wave_stiffness}) can be expanded in powers of the vortex fugacity $y=\exp(-\Ecore/T)$, which is small at low temperatures. The result is
\begin{equation}
\label{K_equation}
\frac{1}{K} = \frac{1}{K_0} + 4 \pi^3 y^2 \int_a^L \frac{\dd r}{a}\left( \frac{r}{a}\right)^{3-2\pi K},
\end{equation}
so that inclusion of vortices has an effect similar to an increase in temperature. Eq.~(\ref{K_equation}) is the basis for a renormalization-group analysis~\cite{Kosterlitz:1974sm, Jose:1977gm, Amit:1979ab} showing that, upon an infinitesimal variation of the lattice spacing $a \to a \exp(\ell) \simeq a (1+\ell)$, the parameters $K$ and $y$ vary as
\begin{equation}
\frac{1}{K} \to \frac{1}{\KR} = \frac{1}{K} + 4 \pi^3 y^2 \int_a^{ae^\ell} \frac{\dd r}{a}\left( \frac{r}{a}\right)^{3-2\pi K}, \qquad y \to \yR = y \exp[(2-\pi K) \ell].
\end{equation}
Finally, these equations can be rewritten in differential form as
\begin{equation}
\ell \frac{d \KR^{-1}}{d \ell} = 4 \pi^3 \yR^2 + O(\yR^4), \qquad \ell \frac{d \yR}{d \ell} = (2-\pi \KR) \yR + O(\yR^3),
\end{equation}
or, setting $\mathcal{X}=2-\pi \KR$ and $\mathcal{Y}=4\pi \yR$ (and neglecting subleading terms),
\begin{equation}
\label{rg_equations}
\ell \frac{d \mathcal{X}}{d \ell} = \mathcal{Y}^2, \qquad \ell \frac{d \mathcal{Y}^2}{d \ell} = 2 \mathcal{X}\mathcal{Y}^2,
\end{equation}
whose solutions near $\mathcal{X}=\mathcal{Y}=0$ are hyperbol{\ae} $\mathcal{X}^2-\mathcal{Y}^2=\mbox{const}$. When the hyperbola vertices lie on the $\mathcal{X}$ axis and the initial value of $\mathcal{X}$ is negative (that is, $T<\TKT$), the renormalization flow drives the system to $\mathcal{Y}=0$ and $\KR$ to a finite value $\KR^\infty$: this corresponds to a line of low-temperature critical points, where correlations decrease as $r^{-T/(2\pi \KR^\infty)}$. On the other hand, in the high-temperature phase one has $\mathcal{X}^2-\mathcal{Y}^2=-s^2<0$; it is then convenient to parametrize $\mathcal{X}$ and $\mathcal{Y}$ in terms of a variable $\psi$, ranging from an initial value $\psiin$ to $\pi/2$, and such that $\mathcal{X}=s \tan \psi$, $\mathcal{Y}= s \sec \psi$, and $\ell = \exp[(\psi -\psiin)/s]$. When the initial value of $\mathcal{X}$ (and thus $\psiin$) is negative, the domain where eqs.~(\ref{rg_equations}) are valid corresponds to $\mathcal{Y} \simeq s$, i.e. to
\begin{equation}
\label{lambda_psiin}
\lambda \simeq \exp (- \psiin/s).
\end{equation}
This value can be interpreted as the largest length scale (in units of $a$) at which the system remains nearly critical, i.e. with the correlation length in units of the lattice spacing. Then, the initial values of $\mathcal{X}$ and $\mathcal{Y}$ must be close to the critical line $\mathcal{Y}=-\mathcal{X}$ (hence $\sin \psiin = \mathcal{X}/\mathcal{Y} \simeq -1$, i.e. $\psiin$ is close to $-\pi/2$). Then, denoting the critical value of $\mathcal{X}$ corresponding to that initial value of $\mathcal{Y}$ as $\Xc$ (i.e. $\Xc=-\mathcal{Y}$), we have $s^2=\mathcal{Y}^2-\mathcal{X}^2=\Xc^2-\mathcal{X}^2=2\Xc (\mathcal{X}-\Xc)$, i.e. $\mathcal{Y} \propto (\mathcal{X}-\Xc)$ or, equivalently, $s^2 \propto (\Kc - K)$. Plugging this result into eq.~(\ref{lambda_psiin}), one eventually finds that, for $\tau=T/\TKT-1 \to 0^+$, the correlation length diverges as described by eq.~(\ref{xi_singularity}), where the constant $b$ is positive.

\end{appendix}

\bibliography{paper}

\end{document}